\documentclass[useAMS,usenatbib]{mn2e}
\usepackage{latexsym,graphicx,natbib,amssymb,array,aas_macros}

\newcommand{\mpc}{M_{\odot}\;{\rm pc}^{-3}}

\newcommand{\kms}{{\rm~km~s}^{-1}}

\title[Universality of multiple star formation]{Revisiting the universality of (multiple) star formation in present-day star formation regions}

\author[M. Marks et. al.] { Michael Marks,$^1$\thanks{e-mail: mmarks@astro.uni-bonn.de (MM)} Nathan Leigh,$^{2,3}$ Mirek Giersz,$^4$ Susanne Pfalzner,$^5$\newauthor Jan Pflamm-Altenburg,$^1$ Seungkyung Oh$^1$\\
$^1$Helmholtz-Institut f\"ur Strahlen- und Kernphysik, University of Bonn, Nussallee 14-16, D-53115 Bonn \\
$^2$Department of Astrophysics, American Museum of Natural History, Central Park West and 79th Street, NY 10024, USA \\
$^3$Department of Physics, University of Alberta, CCIS 4-183, Edmonton, AB T6G 2E1, Canada \\
$^4$Nicolaus Copernicus Astronomical Centre, Polish Academy of Sciences, ul. Bartycka 18, 00-716 Warsaw, Poland \\
$^5$Max-Planck-Institut f\"ur Radioastronomie, Auf dem H\"ugel 69, D-53121 Bonn, Germany }

\begin{document}

\date{Accepted ????. Received ?????; in original form ?????}

\pagerange{\pageref{firstpage}--\pageref{lastpage}} \pubyear{2014}

\maketitle

\label{firstpage}

\begin{abstract}
Populations of multiple stars inside clustered regions are known to change through dynamical interactions. The efficiency of binary disruption is thought to be determined by stellar density. King and collaborators recently investigated the multiplicity properties in young star-forming regions and in the Galactic field. They concluded that stellar density-dependent modification of a universal initial binary population (the standard or null hypothesis model) cannot explain the observations. We re-visit their results, analysing the data within the framework of different model assumptions, namely non-universality without dynamical modification and universality with dynamics. We illustrate that the standard model does account for all known populations if regions were significantly denser in the past. Some of the effects of using present-day cluster properties as proxies for their past values are emphasized and that the degeneracy between age and density of a star-forming region cannot be omitted when interpreting multiplicity data. A new analysis of the Corona Australis region is performed within the standard model. It is found that this region is likely as unevolved as Taurus and an initial density of $\approx190\mpc$ is required to produce the presently observed binary population, which is close to its present-day density.
\end{abstract}

\begin{keywords} methods: data analysis -- methods: statistical -- binaries: general -- stars: formation -- stars: kinematics and dynamics. \end{keywords}

\section{On the universality of star formation}
\label{sec:intro} To within the observational uncertainties, star formation has been observed to be remarkably self-similar in the present-day Universe. Although theory suggests dependences of the initial distribution of stellar masses on, e.g., the metallicity \citep[affecting the Jeans mass;][]{l1998,bcb2006} and/or density \citep{bbz1998,s2004} of a star-forming region, the stellar initial mass function (IMF) is observed to be indistinguishable from the canonical \citet{k2001} IMF in present-day star-forming events \citep[see reviews by][]{bcm2010rev,k2013rev}. In particular, observations of the seven young star-forming regions discussed in the following have previously been shown to be consistent with the universal canonical IMF \citep[see][]{cnk2000cha,b2001oph,m2002onc,p2002usco,t2002cra,p2003ic348,tk2008tau}.

The formation of stars in binary systems is a dominant channel of star formation \citep[see the recent reviews by][]{dk2013rev,reipurth2014rev}. In a stellar population the multiplicity fraction, MuF, of stars which are binaries varies rather strongly between regions. Values of the MuF may reach up to a global frequency of $\approx90$ per cent for sparse, young regions like Taurus \citep{kl1998,d1999} and may be lower than $\approx10$ per cent in some dense, old globular clusters \citep{s2007,m2012acs}. It has been demonstrated that dynamics play an important role in modifying a binary population and cannot be ignored when trying to constrain the initial conditions of star formation \citep{k95a,k95b,gs2000,pz2001,f2003,i2005,thh2007,s2008,fir2009,pgkk2009,kaczmarek2011,mko11,pg2012,leigh2013}. In particular the initial MuF has been argued to be close to $100$ per cent \citep[e.g.][]{i2005,reipurth2014rev} and a universal initial binary population (IBP) has been suggested \citep{k95b}.

Even in the pre-stellar phase, \citet{connelley2008proto} show observationally that multiplicity properties evolve on very short time-scales. The smoothed particle hydrodynamics (SPH) computations of \citet{bate2009a,bate2012sph} demonstrate that gas-induced alterations of orbital parameters and dynamical processing through encounters are non-negligible during cluster formation. \citet{stahler2010} describes analytically the orbital-decay and possible merger of embedded binary stars. Building on such observations \citet{kkp2012} highlight the importance of gas-induced dynamics by demonstrating how it can significantly deplete a binary population for low separations before stars have reached the main-sequence. We note that such processes are included in the \citet{k95b} IBP used here. In this model a semi-analytic description of what is termed 'pre-main-sequence eigenevolution' treats the effects of orbit shrinking and tidal circularization due to dissipation of energy and angular momentum as well as the 
possible mass transfer between and merging of binary components.

In contrast, the environment in which binaries form is probably not universal. It ranges from low-density regions like Taurus ($\approx6$~stars pc$^{-3}$) to dense star-forming clusters in the Galactic Centre \citep[$\approx10^5\mpc$;][]{figer1999arches}. In this context, if the universal IBP is placed inside environments of different density, the observed differences in the present-day multiple star content could be attributed to density-dependent dynamical modification very early on in the cluster lifetime.

A universality of the IBP in itself would not be surprising if the IMF is also universal. Through pre-main-sequence eigenevolution the formation of multiples will influence the resulting distribution of stellar masses and can thus not
easily be separated from the IMF. This notion led to the formulation of the hypothesis that the IBP ought to be invariant if the IMF is \citep{k2011,kp2011}. If this hypothesis were true, the detection of a significant deviation from standard IMF shapes in one region would then also question the universality of multiple star formation in this region.

However, recently \citet*[in the following K12b]{King2012b} claim to find evidence for non-universal (multiple) star formation when comparing Taurus (Tau), Upper Scorpius (USco), Corona Australis (CrA), Chamaeleon (Cha), Ophiuchus (Oph), IC~348, the Orion nebula Cluster (ONC) and the Milky Way field. Assuming that densities have not been significantly denser in the past, K12b analyse the observed multiplicity fractions and separation distributions as a function of the present-day stellar density in these young star formation regions. They find that the observed separation distributions are largely indistinguishable, in contrast to \citet{kp2011} who say that not all present-day populations can stem from a common parent population. From this, K12b conclude that stellar-density-dependent dynamical modification of a universal IBP (referred to as the 'standard model' by K12b) is an incorrect description of the observations. Previous work has however already demonstrated that these very same regions are 
consistent with the standard model, but requires significantly larger initial densities \citep{mk12}.

In this work, we investigate whether the presently available data distinguish differences in the observed orbital separation distributions and how statistically significant they are. In Section~\ref{sec:bdfs}, we demonstrate that the result of K12b's statistical tests meets the expectations of the standard model. Different possible interpretations of K12b's results are discussed in Section~\ref{sec:assumptions}. We close with a discussion of the observational evidence supporting the universality hypothesis of multiple star formation in Section~\ref{sec:universal} and a summary in Section~\ref{sec:summary}.

\section{Statistically analysing observed orbital-parameter distributions}
\label{sec:bdfs}
\begin{figure*}
\includegraphics[width=0.45\textwidth]{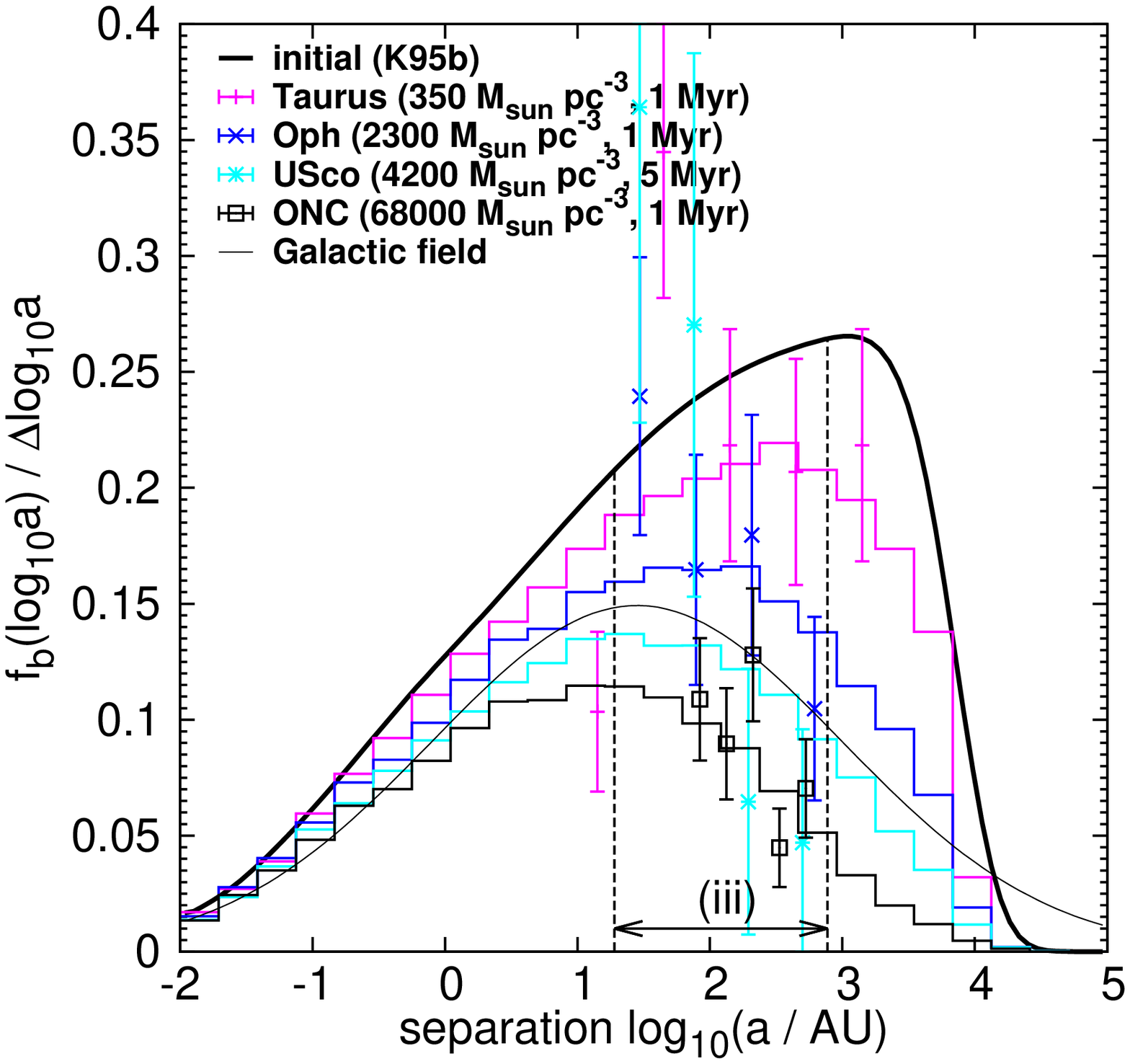}
\includegraphics[width=0.45\textwidth]{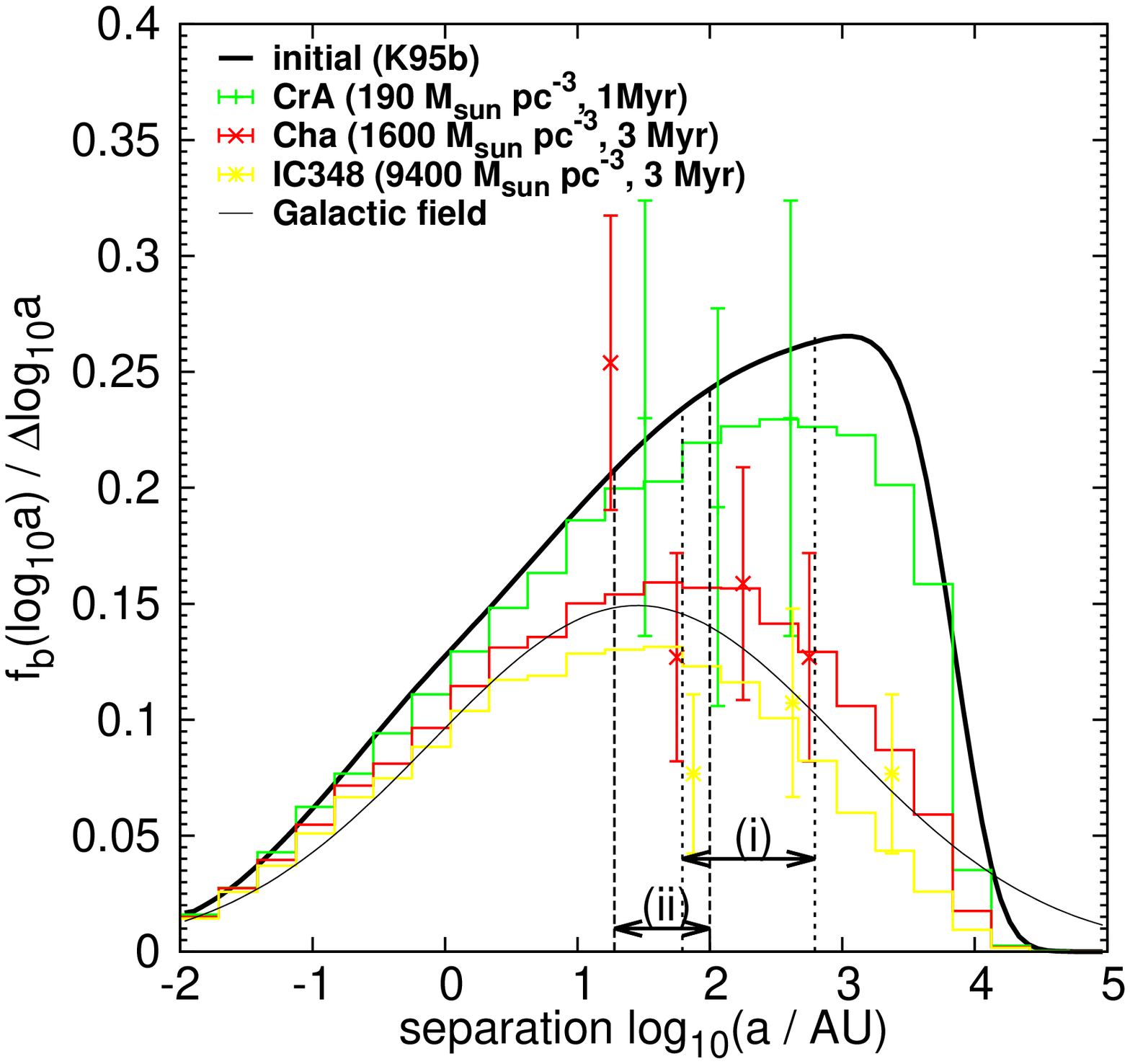}
\caption{Observed separation distributions (points with errorbars) for Cha \citep{l2008cha}, CrA \citep{kn2008cra}, Oph \citep{r2005oph}, Tau \citep{lz1993tau,kl1998}, USco \citep{k2000usco}, IC348 \citep{dbs1999ic} and the ONC \citep{r2007onc} that have been analysed both in K12b and \citet{mk12}. CrA has additionally been analysed for this contribution. Overplotted as the thick solid line is the initial binary distribution suggested by \citet{k95b}. The thin solid line is the field separation distribution for solar-type stars \citep{DuqMay91}. Theoretical binary separation distributions as derived from dynamical processing of the \citet{k95b} initial separation distribution function that match the corresponding observational data are shown as histograms \citep{mk12}. The ranges between the dashed lines labelled group~(i) ($62-620$~au) and~(ii) ($19-100$~au) are the separation ranges referred to as soft and hard, respectively, by K12b. The dashed range labelled group~(iii) ($19-774$~au) is the full 
separation range. The numbers in parentheses in the key are the necessary initial stellar densities and the (observationally constrained) age to evolve the initial distribution into the best-matching separation distribution functions for the observational data.} \label{fig:obstheodata}
\end{figure*}
A binary orbital-parameter distribution is constructed from binned data and depicted in diagrams of the MuF (or variants thereof) versus the parameter of interest. In Fig.~\ref{fig:obstheodata}, the observed separation distributions for the star formation regions discussed here are depicted. In order to decide whether two such separation distributions are 'similar' the aim is to investigate whether they may stem from similar 'parent distributions' using an objective test.

\subsection{Tests}
\begin{figure}
\includegraphics[width=0.22\textwidth]{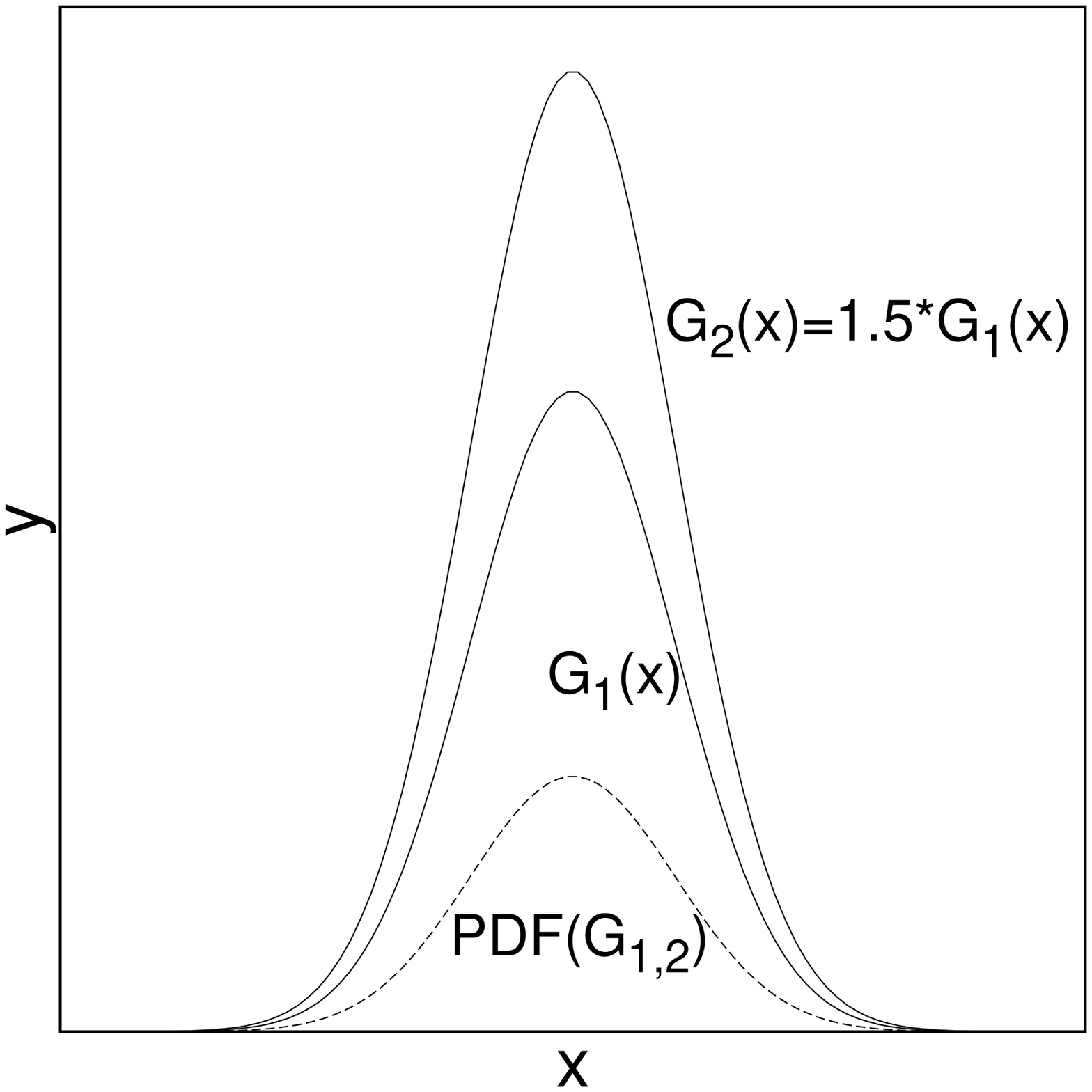}
\includegraphics[width=0.22\textwidth]{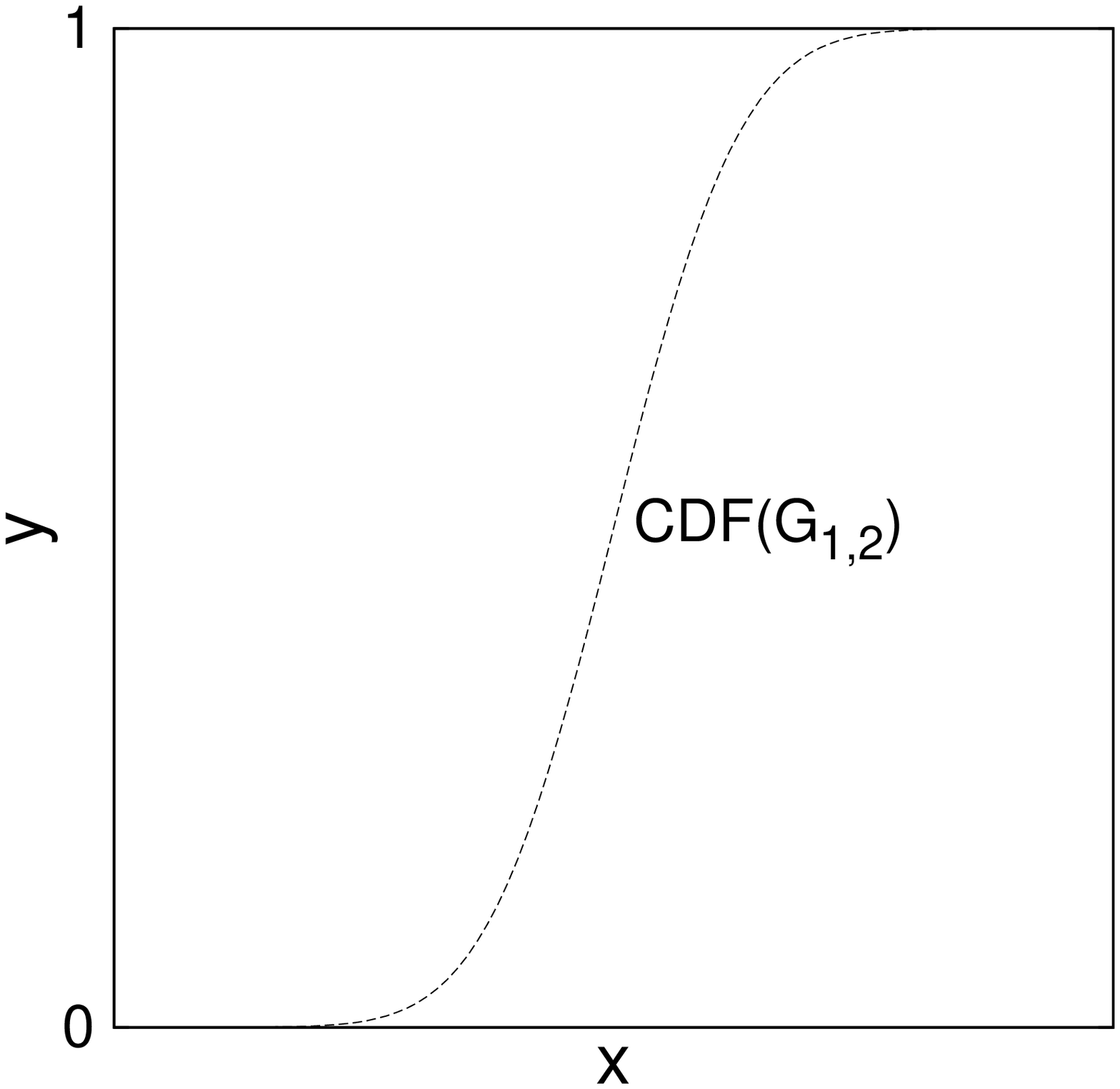}
\caption{Left: two Gaussian (orbital-parameter) distributions $G_1(x)$ and $G_2(x)=1.5G_1(x)$ (solid curves) with the same mean and dispersion share the same probability density function (PDF, normalized such that $\int^{\infty}_{-\infty} G_{1,2}(x)\;dx=1$, dashed curve). Right: the cumulative distribution function (CDF) for the two Gaussian distributions is constructed from their mutual PDF, i.e. both $G_1(x)$ and $G_2(x)$ have the same CDF.} \label{fig:ks}
\end{figure}
One variant that circumvents the need for binning data is the Kolmogorov--Smirnov (KS) test tailored to determine differences or similarities among \emph{probability density functions}. In terms of binary populations these probability density functions are orbital-parameter distributions normalized to unity. Using a KS test differences between two distributions are detected only if their \emph{shapes} are sufficiently different. Two orbital-parameter distributions which differ in the MuF but not in shape would thus remain undiscovered by a KS test (see Fig.~\ref{fig:ks}).

A $\chi^2$-test in contrast requires binned data. The advantage over a KS test is, however, that it detects differences in the MuF even if two distributions are of similar shape. In other words, the difference between a KS and a $\chi^2$-test is that the 'parent distribution' to test for in the former is the underlying probability density function while for the latter it is the actual orbital-parameter distribution.\footnote{Note, though, that a $\chi^2$-test can also be used to distinguish
probability density functions.}

\subsection{Distinguishability of separation distributions}
K12b apply the KS test to the observed separation distributions in three separation ranges (see Fig.~\ref{fig:obstheodata}). They find most of the distributions to show 'no statistically significant differences'. More precisely, however, it is the \emph{shapes} of the distributions that are indistinguishable. This is not surprising given the similar shape among the observed distributions (Fig.~\ref{fig:obstheodata}). Information about the MuF is lost upon (intrinsically) normalizing the separation distributions when applying a KS test. When K12b discuss binned separation distributions, they are thus actually referring to the underlying probability density function. This is evident from their fig. 3 where they compare a cumulative with a binned separation distribution.

The regions Tau, Cha, USco and the ONC also analysed by K12b as well as the Galactic field population have before been statistically analysed by \citet{kp2011}. Using a $\chi^2$-test for their analysis instead, they excluded the hypothesis that the canonical solar-type field separation distribution of \citet{DuqMay91} is similar to the one in Tau with high significance. This instead suggests 'field-like' and 'Tau-like' distributions to have different parent distributions. The remaining tests performed by \citet{kp2011} compared observations against potential parent distributions rather than observations among each other.

In order to perform a reasonable comparison with the K12b analysis, we applied a $\chi^2$-test\footnote{As described in \citet{kp2011}.} to the same observational data as used by K12b, i.e. imposing the same restrictions on separations, contrast limit and mass range.\footnote{The required data are provided by S.~Goodwin.}

Since K12b have separately demonstrated that both the shape and the multiplicity fractions in their restricted samples are similar, most of the results from the $\chi^2$-test suggest the majority of tests not to be able to distinguish the respective separation distributions with more than $90$ per cent confidence, although the detection rate is somewhat larger compared to the KS test.

In group~(i) the null hypothesis that two distributions are similar can be discarded for Tau and ONC with $99.5$ per cent confidence, with $97$ per cent for Tau and Oph and with $95$ per cent for Tau and Cham. Tau and CrA are different in group~(ii) with $99.5$ per cent confidence, Oph and CrA with $97.5$ per cent and Tau and Cha with $93.5$ per cent. Tau and Oph in group~(iii) are different from one another with $99.5$ per cent confidence, Tau and Cha with $96.5$ per cent, Tau and CrA with $93$ per cent and Oph and USco with $91$ per cent.

For comparison, K12b find Oph and USco in group~(i) to be different with $92$ per cent confidence, and that Tau is different from both Cha (with $97.5$ per cent confidence) and CrA (with $92.7$ per cent confidence) in group~(ii). In group~(iii) no separation distribution can be said to be different from one another with more than $90$ per cent confidence according to the KS test.

Whether or not this indistinguishability of most region's separation distributions is a surprising result within the standard model is the subject of investigation in the next section.

\subsection{Comparing separation distributions using statistical tests: a Monte Carlo approach}
\label{sec:reanalyze}
\begin{figure}
\includegraphics[angle=270,width=0.45\textwidth]{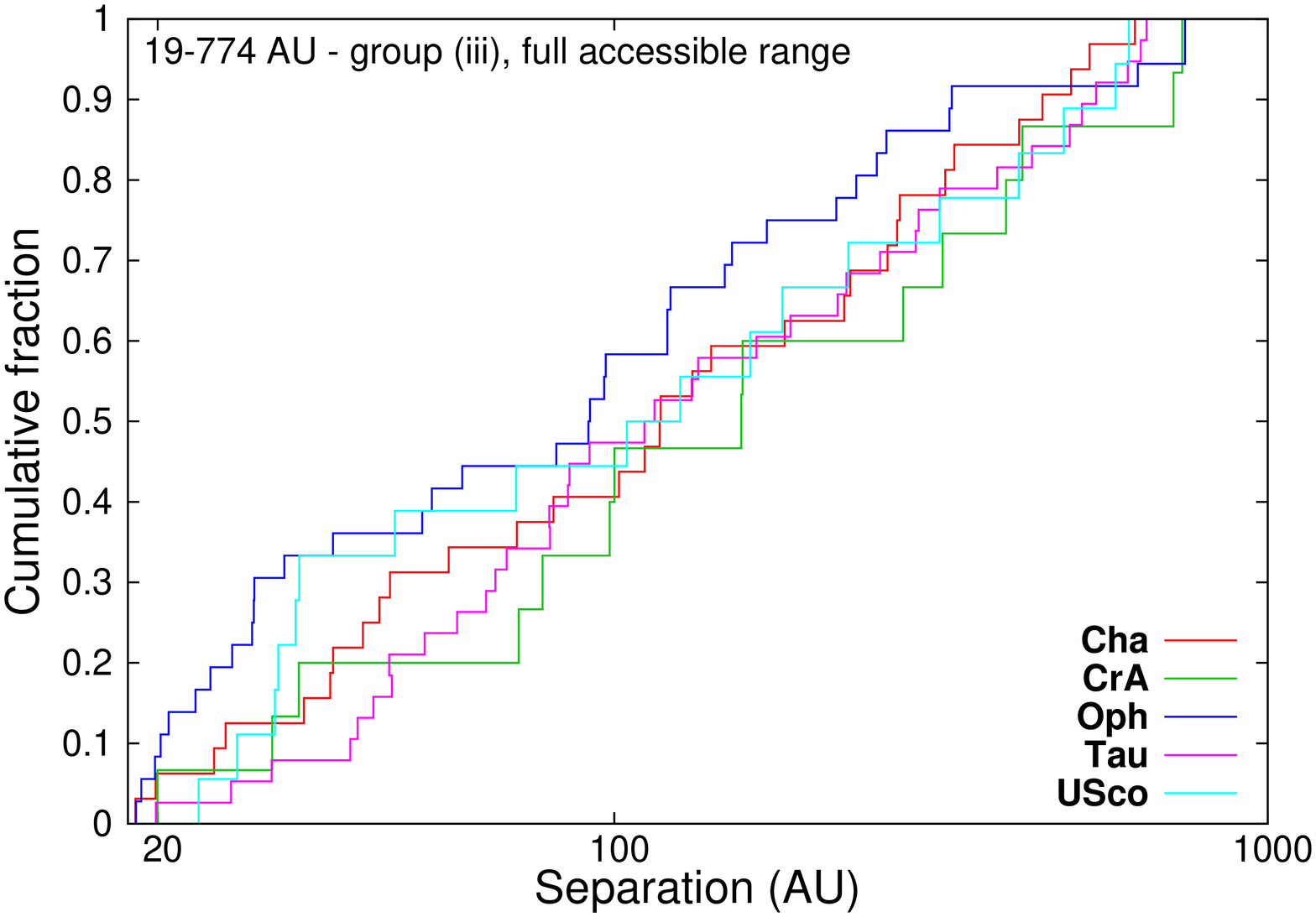} \\
\includegraphics[angle=270,width=0.45\textwidth]{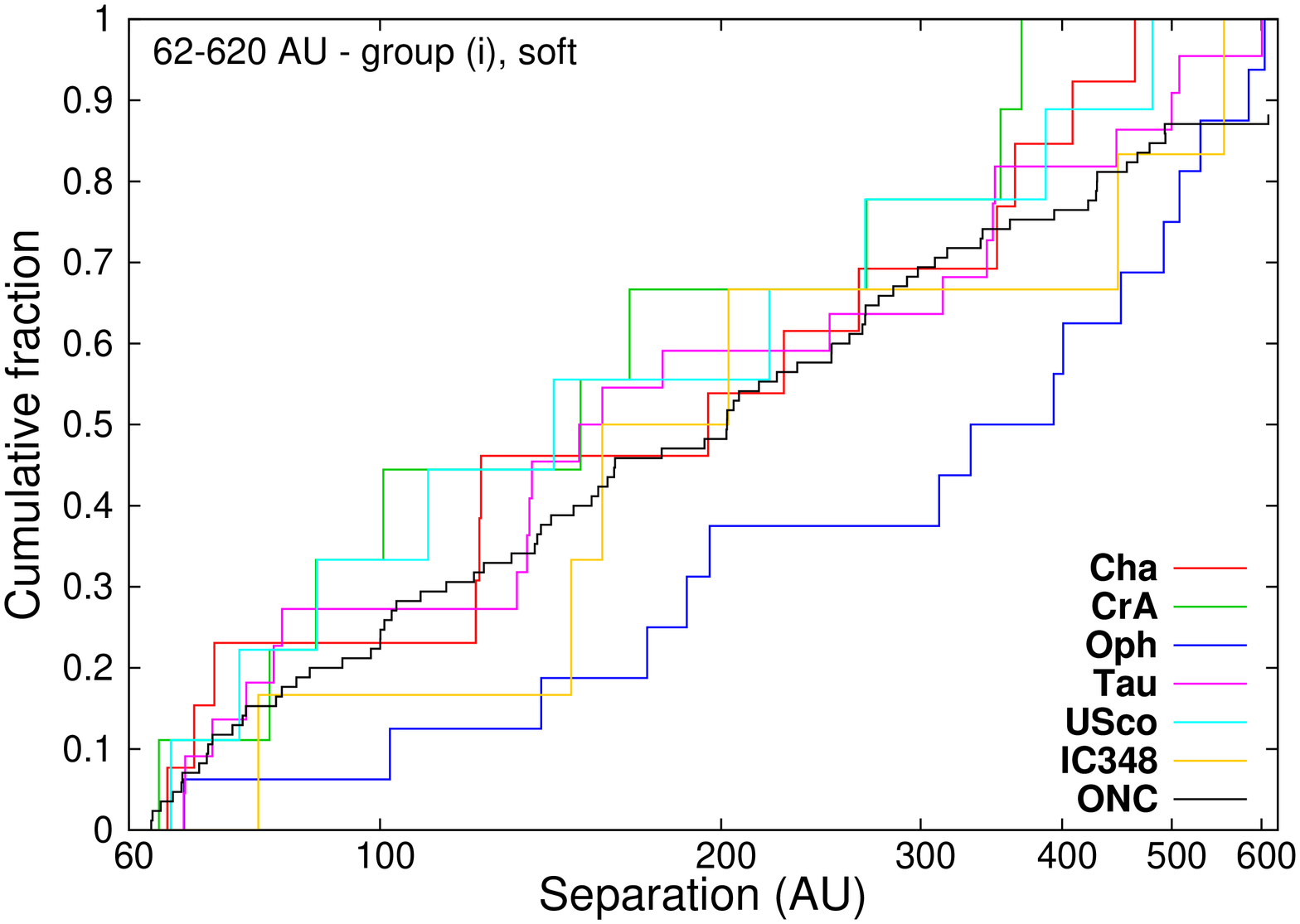} \\
\includegraphics[angle=270,width=0.45\textwidth]{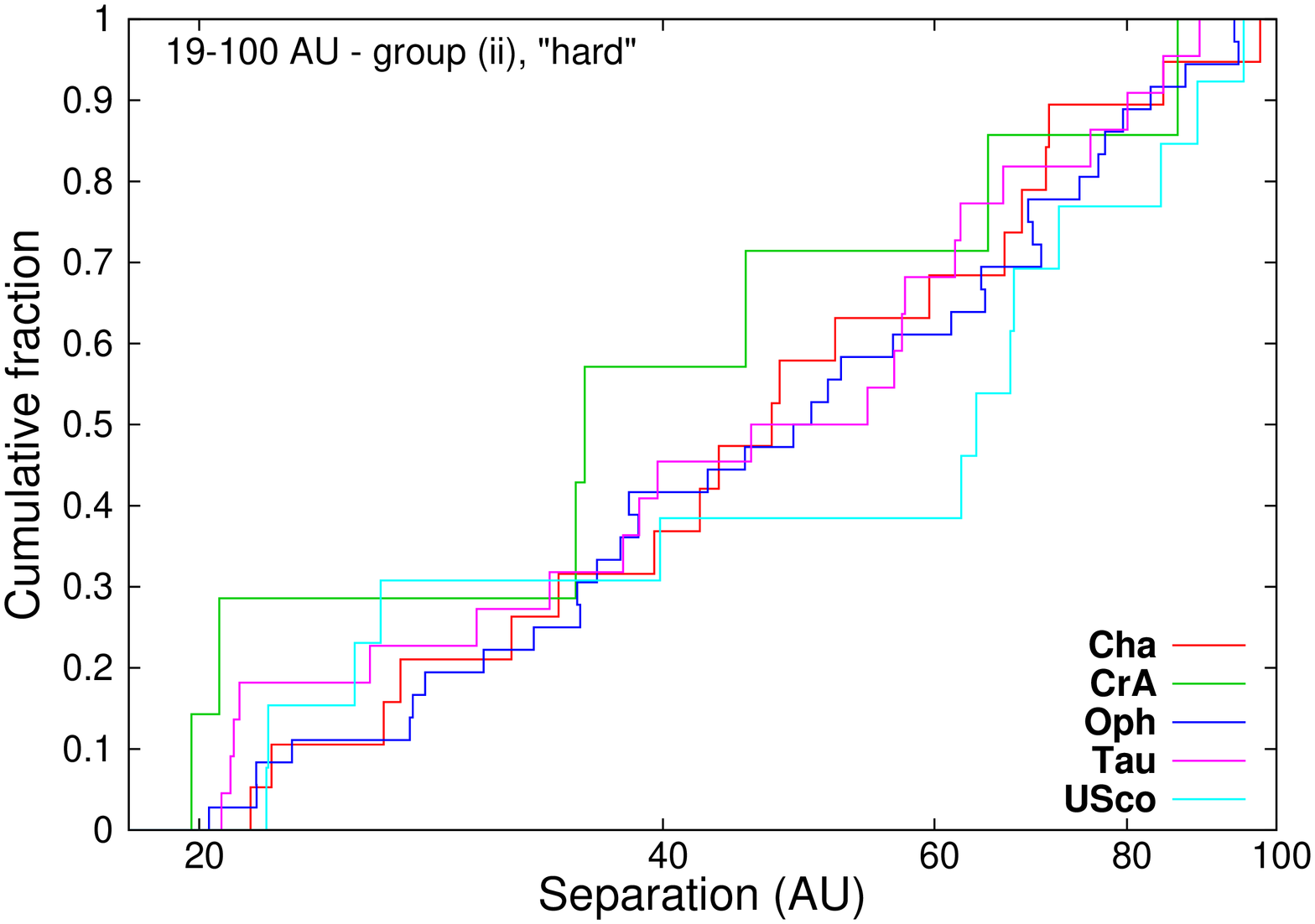} \\
\caption{Cumulative KS distributions as in fig. 5 of K12b but generated from one random realization of each theoretical separation distribution (Fig.~\ref{fig:obstheodata}) that matches observed data of a particular star formation region in the standard model.} \label{fig:kscumdist}
\end{figure}
The conclusion of K12b that the standard model does not account for the observations can only be drawn if the separation distributions are altered enough, and the number of detected binaries is sufficiently large for differences to be detected by a KS or $\chi^2$-test. Otherwise, dynamics-induced differences in the separation distributions would possibly remain undetected. \citet{mk12} tested the standard model against largely \emph{the same} observational data as K12b.\footnote{In contrast to K12b no restrictions on the data were made. This was not necessary as the data were compared against the standard model and not among each other.} They found that for each region a solution exists within the standard model if the initial densities were significantly larger than today. The obtained separation distributions are probable parent functions for the observational data.

\citet{mk12}'s results are overplotted for comparison with the observational data in Fig.~\ref{fig:obstheodata} (histograms). Notably, the model separation distributions are mostly bell-shaped. Within the standard model a dense region will have passed the states of less dense regions (e.g., the ONC will subsequently have had the current separation distributions of Tau, Oph and USco for a short while in the past; left-hand panel in Fig.~\ref{fig:obstheodata}). From this it can be seen that the widest binaries are preferentially broken apart. The key point to take away is that the peak of the separation distribution migrates to lower separations due to dynamical processing, but the overall distribution shape is roughly preserved.

\begin{table*}
\begin{center}
\caption{Groups of separation ranges defined by K12b (columns 1 and 2) and the lowest KS-probabilities for the comparison of two regions for the observations (column 3) and one particular random realization of the theoretically evolved IBP for each region (column 4, the same realization as in Fig.~\ref{fig:kscumdist}). The last two columns give the fraction of comparisons for which the null hypothesis 'Two distributions have the same parent distribution' were discarded if an arbitrary level of $10$~per~cent is imposed (column 5 for the observations, column 6 for $1000$ MC realizations of the dynamically processed IBP for each region). The numbers in parentheses state the total number of comparisons performed (columns 3 and 4) and the fraction of comparisons for which the null hypothesis was discarded (columns 5 and 6).}
\begin{tabular}{>{\footnotesize }c>{\footnotesize }c>{\footnotesize }c>{\footnotesize }c>{\footnotesize }c>{\footnotesize }c}
\hline
& & \multicolumn{2}{c}{Lowest KS probability} & \multicolumn{2}{c}{Null hypothesis discarded at $10$ per cent level}\\
Group & Sep. range (au) & Observations & Processed IBP & Observations & Processed IBP\\
\hline
(i) & 62-620 & $8$ per cent ($21$) & $0.7$ per cent, $4$ per cent ($21$) & $5$ per cent (1/21) & $24$ per cent (16k/21k) \\
(ii) & 19-100 & $2.5$ per cent, $0.63$ per cent ($10$) & $12$ per cent ($10$) & $20$ per cent (2/10) & $14$ per cent (8.6k/10k) \\
(iii) & 19-774 & $15$ per cent ($10$) & $9$ per cent ($10$) & $0$ per cent (0/10) & $12$ per cent (11.2k/10k) \\
\hline
\end{tabular}
\label{tab:ks}
\end{center}
\end{table*}
As an example, in this section it is assumed that this model provides the true underlying parent separation distributions for the observational data. We investigate what results are obtained if KS and $\chi^2$-test are applied to these distributions. CrA's separation distribution was not analysed by \citet{mk12} but this has been done here in the same way. It is found that according to the standard model CrA is as unevolved as Tau and an initial model with stellar density $\rho=190\mpc$ for its present age of $\approx1$~Myr is required to best match the observed separation distribution.

Random realizations of the theoretical distributions in the same three separation ranges as defined by K12b (see Table~\ref{tab:ks}) are produced as a first step. To do so, the probability density functions are constructed from the theoretical separation distributions in the respective groups. Then, a number of separations according to the number of binaries actually observed (see table~2 in K12b) are randomly selected.

Due to the nature of the random selection experiment it cannot be expected that with any realization the observational results are exactly reproduced, neither for the KS or $\chi^2$-probability values, nor that the lowest probability occurs for the comparison of the same two regions. For improved statistical significance, $1000$ realizations are drawn for each region from the theoretical separation distributions and the statistical tests are applied to each.

\subsubsection{KS test}
The cumulative separation distributions generated from the randomly drawn separations are compaired pairwise to calculate the probability that the null hypothesis 'Two regions share similar probability density functions' be true. Examples of KS distributions from these random realizations are depicted in Fig.~\ref{fig:kscumdist}.

A comparison by eye suggests that the cumulative distributions in each panel of Fig.~\ref{fig:kscumdist} look similar to each other, despite originating from the standard model. This is the same notion arrived at by K12b for the cumulative distributions constructed from the observed data. The lowest KS probabilities for the comparison between two regions are reported in Table~\ref{tab:ks} (columns 3 and 4) for these particular random realizations of each region, as well as for the observations. The values are rather similar (but occur between different regions).

Assuming that the null hypothesis is discarded when the KS probability is lower than $10$ per cent, it is counted how often the KS test delivers probabilities below $0.1$ among all comparisons. The results are reported and compared to the actual observations in column 5 and 6 of Table~\ref{tab:ks}.

The majority of all comparisons suggest that the null hypothesis cannot be discarded. Thus, the KS tests for the random realizations of the theoretical separation distributions stemming from a standard model produce the same results as the KS test applied to the observational data.

\begin{table*}
\begin{center}
\caption{For the KS test, the table shows how often a particular comparison suggests the null hypothesis that two distributions are similar can be withdrawn at the $10$~per~cent level for $1000$ random realizations ('detection probability'). The shown percentages vary at most $\approx5$~per~cent in repeated MC experiments. Columns captioned 'obs' show which comparisons among the actual observations could (\checkmark) and could not ({\sffamily X}) withdraw the null hypothesis.}
\begin{tabular}{>{\footnotesize }c>{\footnotesize }c>{\footnotesize }c>{\footnotesize }c>{\footnotesize }c>{\footnotesize }c>{\footnotesize }c>{\footnotesize }c}
\hline
 & & \multicolumn{6}{c}{Null hypothesis discarded at $10$ per cent level}\\
 \multicolumn{2}{l}{Compared regions} & \multicolumn{2}{c}{Group (i)} & \multicolumn{2}{c}{Group~(ii)} & \multicolumn{2}{c}{Group~(iii)} \\
 & & MC & obs & MC & obs & MC & obs\\
\hline
 Cha & CrA & 16 per cent & \sffamily X & 21 per cent & \sffamily X  & 16 per cent & \sffamily X \\
 Cha & IC348 & 18 per cent & \sffamily X  & -- & --  & -- & -- \\
 Cha & ONC & 29 per cent & \sffamily X  & -- & --  & -- & -- \\
 Cha & Oph & 12 per cent & \sffamily X  & 10 per cent & \sffamily X  & 10 per cent & \sffamily X \\
 Cha & Tau & 9 per cent & \sffamily X  & 9 per cent & \checkmark & 10 per cent & \sffamily X \\
 Cha & USco & 14 per cent & \sffamily X  & 10 per cent & \sffamily X  & 8 per cent & \sffamily X \\
 CrA & IC348 & 25 per cent & \sffamily X  & -- & --  & -- & -- \\
 CrA & ONC & 44 per cent & \sffamily X  & -- & -- & -- & -- \\
 CrA & Oph & 16 per cent & \sffamily X  & 22 per cent & \sffamily X  & 15 per cent & \sffamily X \\
 CrA & Tau & 15 per cent & \sffamily X  & 22 per cent & \checkmark & 13 per cent & \sffamily X \\
 CrA & USco & 11 per cent & \sffamily X  & 15 per cent & \sffamily X  & 14 per cent & \sffamily X \\
 IC348 & ONC & 42 per cent & \sffamily X  & -- & -- & -- & -- \\
 IC348 & Oph & 20 per cent & \sffamily X  & -- & --  & -- & -- \\
 IC348 & Tau & 21 per cent & \sffamily X  & -- & --  & -- & -- \\
 IC348 & USco & 28 per cent & \sffamily X  & -- & --  & -- & -- \\
 ONC & Oph & 28 per cent & \sffamily X  & -- & -- & -- & -- \\
 ONC & Tau & 33 per cent & \sffamily X & -- & --  & -- & -- \\
 ONC & USco & 35 per cent & \sffamily X  & -- & --  & -- & -- \\
 Oph & Tau & 10 per cent & \sffamily X  & 7 per cent & \sffamily X  & 11 per cent & \checkmark \\
 Oph & USco & 16 per cent & \checkmark & 10 per cent & \sffamily X  & 9 per cent & \sffamily X \\
 Tau & USco & 14 per cent & \sffamily X  & 9 per cent & \sffamily X  & 8 per cent & \sffamily X \\
\hline
\end{tabular}
\label{tab:ksmc}
\end{center}
\end{table*}
A different kind of comparison is provided in Table~\ref{tab:ksmc}. For each individual comparison of regions it is shown how often among $1000$ random realizations the Monte Carlo (MC) experiment could discard the null hypothesis at the $10$ per cent level if the data are selected from the theoretical distributions. This can be interpreted as the probability for differences between the involved separation distributions to be detected if the data stemmed from the standard model. The results are as follows.
\begin{itemize}
\item[(i)] \textbf{Group (i)}: there are about 10 comparisons with a detection probability between 10-20 per cent. We would thus expect that for the actual observations, one to two comparisons allow the null hypothesis to be withdrawn with at least $90$ per cent confidence. One is observed (tick marks in the table). About six comparisons detect the two involved distributions to be different with $\approx30$ per cent probability. This suggests two comparisons to detect differences among the observed distributions. None is observed. The difference between CrA and the ONC's separation distribution may or may not be detected according to the MC experiment. It is not.
\item[(ii)] \textbf{Group (ii)}: about seven comparisons withdraw the null hypothesis with a detection probability of $\approx$10 per cent each, i.e. zero to one comparisons among the observations should actually distinguish their respective separation distributions. It is the case for one. For the remaining three comparisons, the probability to detect a difference among the separation distributions is about $20$ per cent. A difference is detected in one test and zero to one are expected.
\item[(iii)] \textbf{Group (iii)}: all 10 comparisons detect a difference with $\approx10$ per cent probability. We thus expect one test to actually withdraw the null hypothesis, and this is the case.
\end{itemize}

\subsubsection{$\chi^2$-test} % \setlength\tabcolsep{4pt}
\begin{table*}
\begin{center}
\caption{Same as Table~\ref{tab:ksmc} but for the $\chi^2$-test.}
\begin{tabular}{>{\footnotesize }c>{\footnotesize }c>{\footnotesize }c>{\footnotesize }c>{\footnotesize }c>{\footnotesize }c>{\footnotesize }c>{\footnotesize }c}
\hline
 & & \multicolumn{6}{c}{Null hypothesis discarded at $10$ per cent level}\\
 \multicolumn{2}{l}{Compared regions} & \multicolumn{2}{c}{Group (i)} & \multicolumn{2}{c}{Group (ii)} & \multicolumn{2}{c}{Group (iii)} \\
 & & MC & obs & MC & obs & MC & obs\\
\hline
 Cha & CrA & 26 per cent & \sffamily X & 35 per cent & \sffamily X & 46 per cent & \sffamily X \\
 Cha & IC348 & 18 per cent & \sffamily X  & -- & --  & -- & -- \\
 Cha & ONC & 17 per cent & \sffamily X  & -- & --  & -- & -- \\
 Cha & Oph & 10 per cent & \sffamily X  & 17 per cent & \sffamily X  & 16 per cent & \sffamily X \\
 Cha & Tau & 43 per cent & \checkmark  & 21 per cent & \checkmark & 27 per cent & \checkmark \\
 Cha & USco & 13 per cent & \sffamily X  & 22 per cent & \sffamily X & 28 per cent & \sffamily X \\
 CrA & IC348 & 29 per cent & \sffamily X  & -- & --  & -- & -- \\
 CrA & ONC & 35 per cent & \sffamily X  & -- & -- & -- & -- \\
 CrA & Oph & 30 per cent & \sffamily X  & 37 per cent & \checkmark & 52 per cent & \sffamily X \\
 CrA & Tau & 24 per cent & \sffamily X  & 39 per cent & \checkmark & 46 per cent & \checkmark \\
 CrA & USco & 19 per cent & \sffamily X  & 26 per cent & \sffamily X & 33 per cent & \sffamily X \\
 IC348 & ONC & 63 per cent & \sffamily X  & -- & -- & -- & -- \\
 IC348 & Oph & 22 per cent & \sffamily X  & -- & --  & -- & -- \\
 IC348 & Tau & 67 per cent & \sffamily X  & -- & --  & -- & -- \\
 IC348 & USco & 12 per cent & \sffamily X  & -- & --  & -- & -- \\
 ONC & Oph & 13 per cent & \sffamily X  & -- & -- & -- & -- \\
 ONC & Tau & 100 per cent & \checkmark  & -- & --  & -- & -- \\
 ONC & USco & 30 per cent & \sffamily X  & -- & --  & -- & -- \\
 Oph & Tau & 47 per cent & \checkmark  & 27 per cent & \sffamily X & 37 per cent & \checkmark \\
 Oph & USco & 16 per cent  & \sffamily X  & 27 per cent & \sffamily X & 31 per cent & \checkmark \\
 Tau & USco & 31 per cent & \sffamily X  & 21 per cent & \sffamily X & 40 per cent & \sffamily X \\
\hline
\end{tabular}
\label{tab:chi2mc}
\end{center}
\end{table*}
In order to perform the $\chi^2$-test, the randomly selected separations are distributed in four bins in groups~(i) and~(ii), and in eight bins in group~(iii). These are then normalized to the number of targets searched for companions and the width of the bins. Binomial errors are assumed. In Table~\ref{tab:chi2mc}, the same comparison as for the KS test is performed.
\begin{itemize}
\item[(i)] \textbf{Group (i)}: the difference between Tau and the ONC is obvious in the standard model and is detected for the actual observations using the $\chi^2$-test. Four comparisons suggest differences to be detected with $40--70$ per cent probability (average $55$ per cent). We thus expect differences for two comparisons to be actually observed. This is the case. Five comparisons detect differences in $\approx30$ per cent of the cases. One to two comparisons among the actual observations should thus identify differences. None is observed. Four to five comparisons discard the null hypothesis in 10-15 per cent of the cases. Zero to One actual comparisons should thus yield $\chi^2$ significance probabilities lower than $10$ per cent. Zero are observed. 
\item[(ii)] \textbf{Group (ii)}: six comparisons discard the null hypothesis at the $10$ per cent level with $\approx30$ per cent probability. We expect thus two comparisons to actually refute the hypothesis that their separation distributions are similar. This is the case. Three comparisons detect differences at the $10$ per cent level in $20$ per cent of the cases. For one comparison among the observations, this is actually detected, and we expect zero to one.
\item[(iii)] \textbf{Group (iii)}: four comparisons detect differences with $30$ per cent chance each. We thus expect a difference to be observed for $\approx1$ comparison, observed are two. For five comparisons, differences should be detected in $\approx40$-$50$ per cent of the cases, i.e. $\approx2$ are expected. Observed is a difference for one.
\end{itemize}

\subsubsection{Dependences on number of observed binaries and number of bins}
We briefly note that the detectability of differences among separation distributions is dependent on both the observed number of binaries and the number of bins. By randomly drawing more binaries than are actually observed in all
regions we tested the effect of using a larger number of binaries.

Increasing the number of binaries better probes the underlying parent distributions. If the standard model provides the true distributions, the detection rate of differences among these distributions is reduced if a KS test is applied since the observed distribution \emph{shapes} are closer to the true ones. In contrast, the detection rate of differences in a $\chi^2$-test is enhanced since differences in the MuF for individual bins are more pronounced.

Increasing the number of bins in the $\chi^2$-test additionally reduces the detectability of differences as fewer binaries per bin enhance the level of noise.

\subsubsection{Conclusion}
The results from the KS and $\chi^2$-test applied to the observations matches the expectations from the standard model. This is because the standard model produces separation distributions which are largely indistinguishable by these tests, given their similar shape and the observed numbers of binaries. Caution is thus needed when interpreting statistical tests. Following K12b, one would need to say that the majority of the theoretical separation distributions are the same and that the standard model is excluded. But this is true only if densities were not significantly different from today.

As the same results are obtained within the standard model, both the conclusions by K12b that their 'analysis strongly suggests that this standard model is incorrect' and 'binary formation is not the same everywhere and therefore star formation is not a single universal process' are suspect. Applying a KS or $\chi^2$-test to the observed separation distributions cannot by themselves confirm or refute the validity of the standard model, even if the results suggest that most distributions are indistinguishable. Indeed, the standard model has already passed the test against the observations \citep{mk12}.

\section{Discussion of further results}
\label{sec:assumptions}
We continue to discuss K12b's additional results. The interpretation of these depend on the assumptions concerning the initial densities of the star-forming regions (Section~\ref{sec:softhard}), whether the region's ages are included in the interpretation (Section~\ref{sec:agedensity}) and whether the star formation regions here discussed are representative of star formation environments in general (Section~\ref{sec:field}). We express our concerns about their choices and, in turn, the interpretation of their results.

\subsection{Which binaries are soft and which are hard?}
\label{sec:softhard}
Using the KS test, K12b find that Tau's separation distribution could be different from both Cha and CrA in group~(ii). According to them, this is unexpected in the standard model as these binaries should not be affected through dynamical interactions.

Whether a binary with a given separation inside a population is considered rather easy ('soft', low binding energy) or difficult ('hard', large binding energy) to break up in the course of dynamical processing depends on 'the maximum density a region reaches/reached that sets the degree of destructiveness of encounters -- not necessarily the current density -- and therefore differences/similarities between regions may be due to past differences/similarities'. Thus, a binary may in principle change from soft to hard, and vice versa, in course of a region's history independent of dynamical interactions altering its orbital parameters.

Stellar densities can evolve even (or especially) in young regions. Mass-loss processes due to stellar evolution \citep{vesperini2009} or residual-gas expulsion \citep{bk07}, as well as the energy generated through binary hardening \citep{mko11} drive star formation regions into a more extended state. All these mechanisms act on rather short time-scales (a few crossing time-scales, i.e. up to a few Myr).\footnote{Later two-body relaxation also tends to increase the central concentration (assuming that the tidal radius stays roughly constant). Relaxation, however, typically is relevant only on a time-scale much longer than the age of these regions.} Observations of molecular cloud clumps at the verge of star formation are observed to be denser than the here discussed star formation regions \citep{mueller2002mc,shirley2003mc,fontani2005mc}. This might be interpreted as observational evidence that the progenitors of star formation regions, i.e. dense molecular clouds, are required to expand to appear in their present shape.

\citet{bate2009a} reports the velocity dispersion for the central cluster formed in his SPH computations to be $\approx4\kms$. When including radiative transfer in his SPH models, \citet{bate2012sph} finds that the overall rms velocity dispersion is $5.5\kms$ in three dimensions and $3.2\kms$ in one dimension. This is larger than the value adopted by K12b by a factor of $\approx$2. Increasing the velocity dispersion by a factor of 2 quarters the location of the hard--soft boundary as
\begin{equation}
a_{\rm hs}=450\left(\frac{\sigma}{\kms}\right)^{-2}\;{\rm au}. \label{eq:hardsoft}
\end{equation}
Therefore, it might be expected that a larger fraction of binaries were soft in the past such that more and lower separation binaries could be dissolved due to encounters within the same time (Section~\ref{sec:agedensity}) than might be expected from adopting present-day values.\footnote{As an aside, it is cautioned that this approach relies on averages and that the break-up of binaries is mass dependent. Whether a binary is actually hard or soft is relative because it depends on the parameters of the individual encounter. For example, a binary being hard according to equation~\ref{eq:hardsoft} may actually be soft in an energetic encounter with another star or binary with a velocity that exceeds the cluster velocity dispersion $\sigma$.}

K12b's adopted velocity dispersion, $\sigma=2\kms$, as an upper limit on its present-day value in Tau and USco \citep{frink1997tau,kh2008usco}, results in $a_{\rm hs}=112.5$~au and leads K12b to define the separation ranges 19-100~au as 'hard' [group~(ii)] and $62-620$~au as 'soft' [group~(i); compare Fig.~\ref{fig:obstheodata}]. \citet{kh2008usco} additionally report observational evidence for small-scale velocity dispersions increasing with clusters density. For the ONC, it is already $\sigma=3.1\kms$ \citep{furesz2008onc}.

Doubling the velocity dispersion would, however, lower the hard--soft boundary to $\approx30$~au. This is only about $10$~au larger than the minimum binary separation used in their analysis and roughly coincides with the peak of the semi-major axis distribution for field binaries \citep{DuqMay91,r2010}. Past dynamical modification of binaries with separations in group~(ii) were then expected. In this case, it is not justified to claim that differences in the observed distributions for group~(ii) were inconsistent with the standard model.

What's more, even if no hard binary in any population was ever disrupted, the MuF of \emph{hard} binaries is lowered through the dissolution of \emph{soft} binaries. Consider the multiplicity fraction \begin{equation} {\rm MuF}=\frac{\rm B+T+Q}{\rm S+B+T+Q}\;, \label{eq:mf} \end{equation} where S, B, T and Q are the numbers of single, binary, triple and quadruple systems, respectively. The MuF decreases in \emph{all} separation ranges through a net increase of the denominator by 1 upon each \emph{soft} binary disruption (in the denominator, B decreases by 1 and S increases by 2, while when considering the MuF of truly hard binaries, the nominator does not change). Thus, if the soft binary MuF in group~(i) is a function of stellar density then the hard binary MuF in group~(ii) is, too, even if the orbital properties of hard binaries were not modified at all since their formation.

Further effects modifying hard binary populations and potentially blurring expectations about primordially hard binaries include the preferential hardening of hard binaries in encounters \citep[the Hills-Heggie law,][]{hills1975,heggie1975}, the break up of hard binaries through stellar evolution \citep{i2005} and ejection of hard binaries from the cluster centre \citep{ok2012,okb2013r144}.

\subsection{The age--density degeneracy} \label{sec:agedensity}
When discussing binary populations in a particular region it is not only important to know its (peak) density but also its age. The older a region the more time there is for binary processing. This leads to a degeneracy between age and density. An initially high density cluster of age $t(1)$ may have the same binary population as a somewhat lower density cluster of age $t(2) > t(1)$ since dynamical processing in the latter cluster is less efficient \citep{mk12}. This can be seen for separation distributions of some regions in Fig.~\ref{fig:obstheodata}, e.g. Oph and Cha have very similar separation distributions albeit having different ages and suggested initial peak densities.

\begin{figure} 
\includegraphics[angle=270,width=0.43\textwidth]{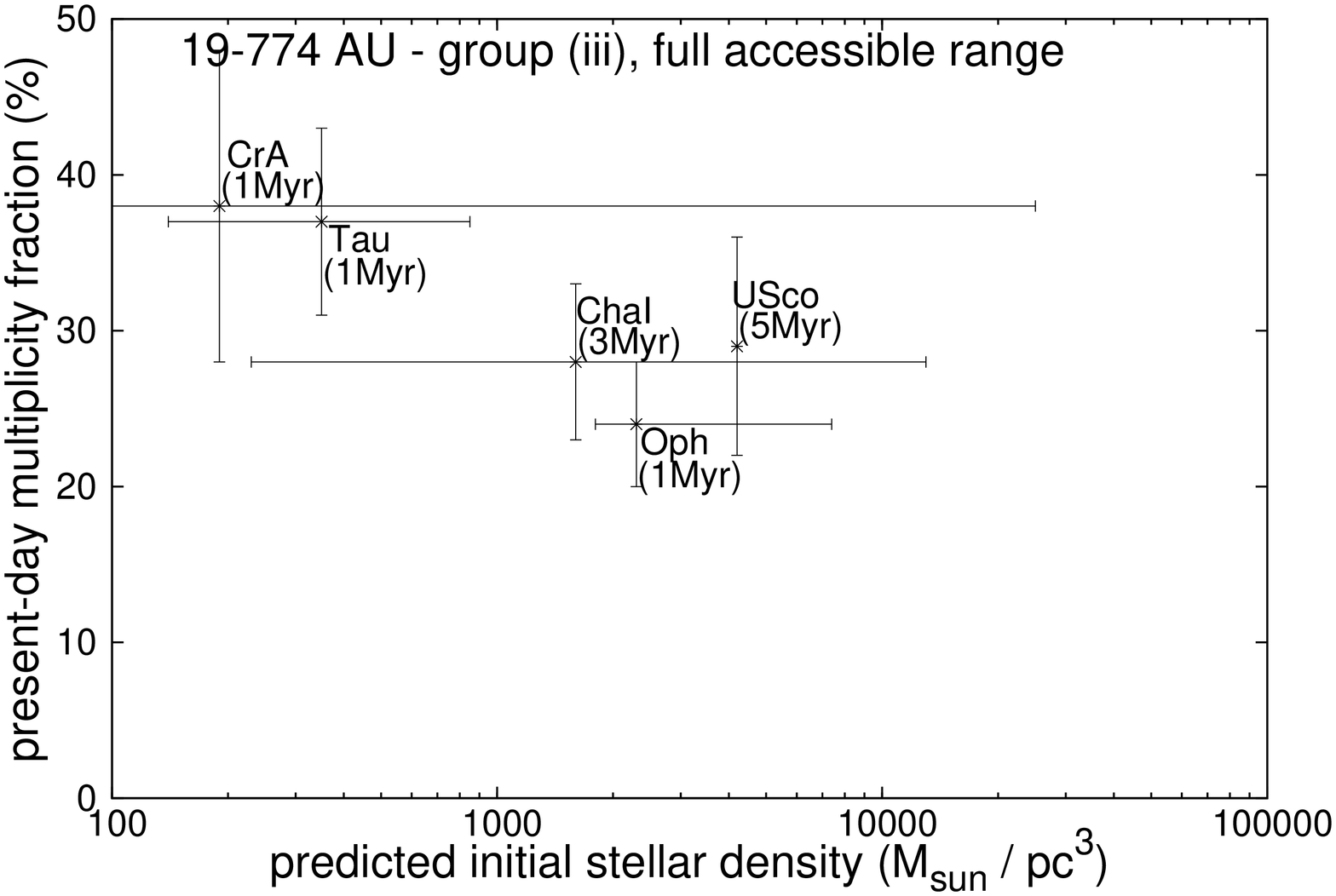} \\
\includegraphics[angle=270,width=0.43\textwidth]{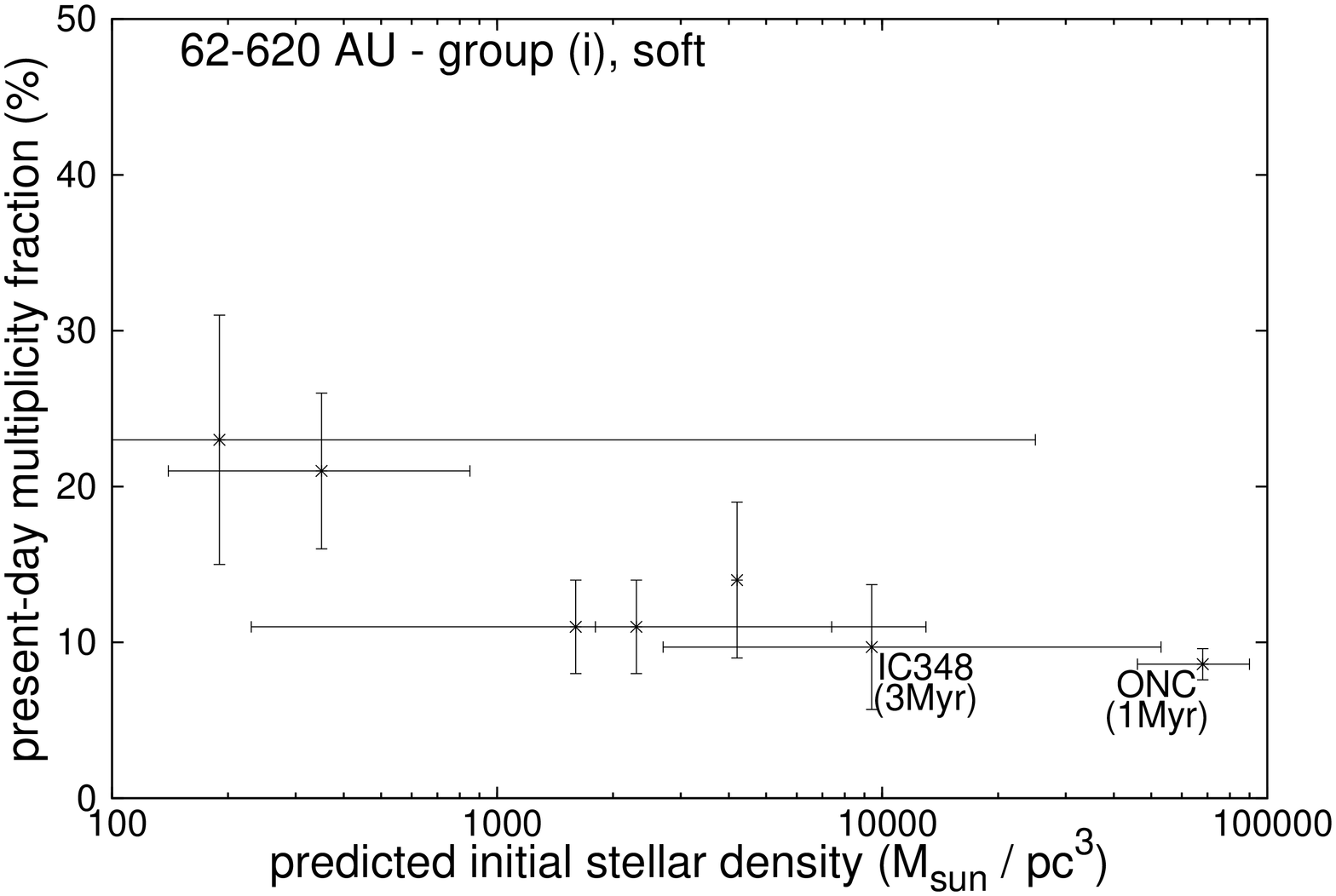} \\
\includegraphics[angle=270,width=0.43\textwidth]{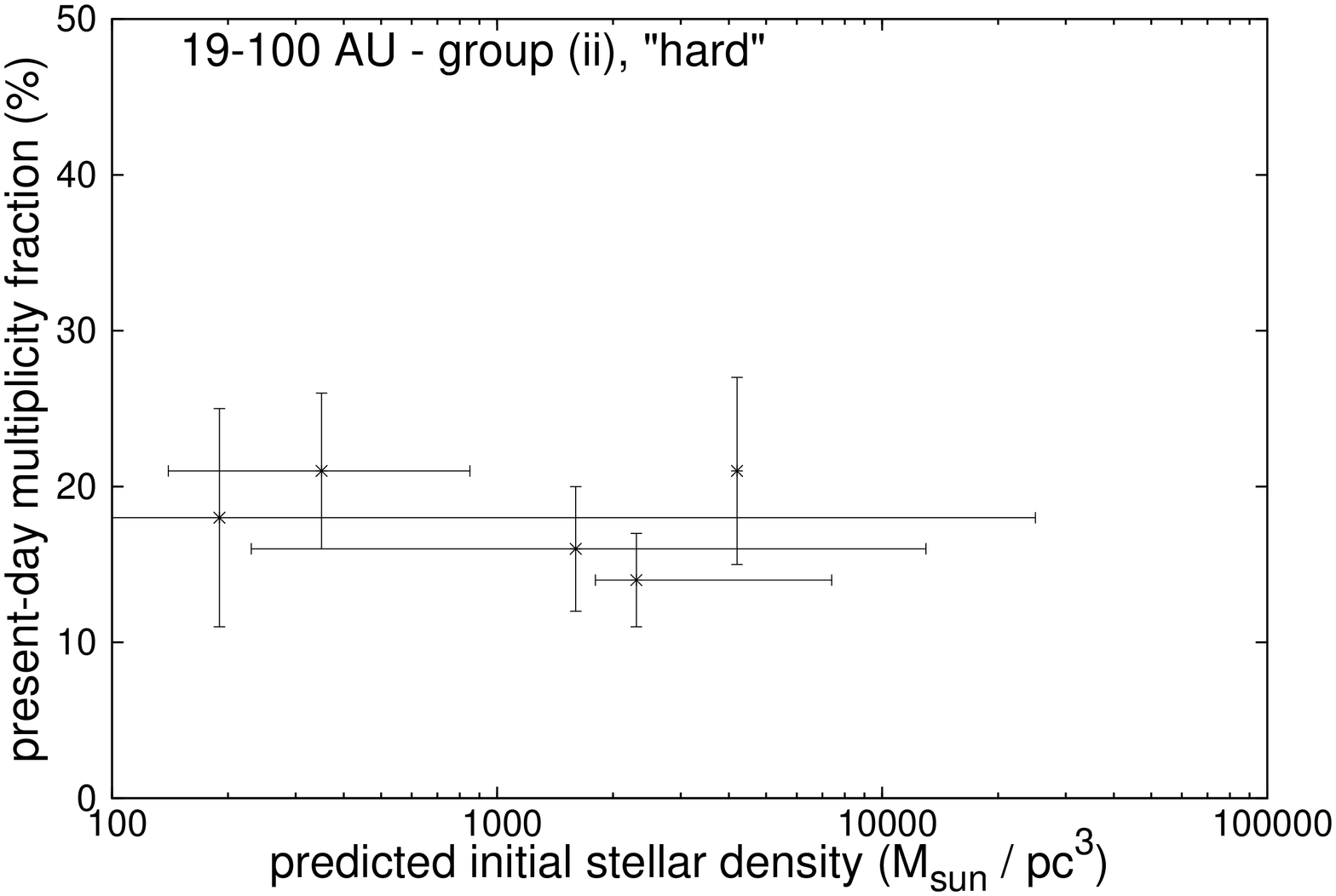} \\
\caption{Present-day binary fractions in groups (i), (ii) and (iii) as a function of the initial stellar density as predicted by \citet{mk12}. Statistical errors for the multiplicity fractions as in K12b. Errors in the initial stellar densities as in
\citet{mk12}. A wide range allowed for the initial densities arises if the observed sample has been small or the bin-to-bin variation of the multiplicity fraction in the available semi-major axis range in the observations is large. For USco, the best-fitting initial density is shown without errorbars as the constraint on its initial density is poor given the observational data (essentially all densities are allowed without discarding the null hypothesis).} \label{fig:binfracs}
\end{figure}
As a function of the present-day density, K12b discuss MuFs within the investigated regions and claim there to be no \emph{significant} correlation of MuF with density. Albeit weak, with the exception of CrA, the data show a correlation of decreasing binary fraction with increasing density in all groups (fig. 4 in K12b). This could be the result of density-dependent processing. In the same sense, \citet{m2012acs} show correlations between the core binary fraction and core densities for globular clusters.

K12b acknowledge this, but note 'that only Taurus and CrA have ann MuF more than 1$\sigma$ above that of the ONC'. The authors go on to say that this is unexpected as USco, Cha, Oph, IC348 and the ONC actually span nearly two orders of magnitude in present-day number density. This can be reconciled by considering the different ages of these regions. USco has an estimated present-day density two orders of magnitude below the ONC, at least an age of about 5~Myr \citep{p2002usco} and may be as old as 11~Myr \citep{pecaut2012usco}. To the present day, it thus had at least five times longer to process its binary population than the ONC. Cha, being more than one order of magnitude below the ONC in terms of the present-day number density, has a median age of 2~Myr \citep{l2004a} and thus had twice the time to break-up its binaries. The outlier CrA is itself a very young region \citep[$\approx1$~Myr,][]{nf2008cra} having a binary fraction more similar to Tau. USco and Cha having about half and double the estimated density of CrA, respectively, but a lower binary fraction are both older than CrA. Thus, taking the observed similar MuF's plus their ages at face value, the only weak correlation between present-day number density and binary fraction is not necessarily a surprising result, even if their corresponding initial densities were not larger in the past.

Fig.~\ref{fig:binfracs} is similar to K12b's fig. 4, but the observed MuF is depicted here as a function of the initial stellar density as constrained by \citet{mk12}. The correlation seen between the MuF and the present-day number density is retained if not enhanced when using the initial density. In this representation the K12b outlier CrA in fact becomes the region with the lowest initial density thus providing an explanation for its large MuF. Also the age-density degeneracy becomes more apparent here. While Cha and Oph have a similar present-day MuF, Cha required a slightly lower density being at least two times older than Oph\footnote{An age of $3$~Myr has been used for Cha in the analysis of \citet{mk12} as an upper limit for its age.} to presently arrive at the same MuF. An outlier in this representation seems to be USco but its initial density is not well constrained given the observational data \citep[see caption of Fig.~\ref{fig:binfracs} and][]{mk12}.

In this model, the initial densities are constrained to be larger than the present-day values by a factor of $\approx10-1000$ in \citet[see their fig.~5]{mk12}.\footnote{The most extreme case reported is the Pleiades (not discussed here) which requires a density $4000$ times its present density.} Whether these are realistic values might be questioned. For example, \citet{wright2014cygob} argue that the massive OB association Cygnus~OB2 is dynamically young based on the existence of substructure and might not have been significantly denser in the past. Cluster half-mass radii in the model would be required to be significantly smaller than observed for any massive cluster today \citep{pmg2010rev}. The quality of the data are reflected in the large errorbars to these constraints. For example, the best model for Taurus suggests it to be at least twenty times as dense as nowadays but it has been shown to be consistent even with the universal initial distribution \citep[i.e. 'zero' initial density;][]{kp2011}. The presented densities are however what current observations allow us to constrain. Rather than the absolute values, the relative differences between the constrained initial densities are possibly a more robust model prediction. As the data quality improves, so will the constraints based on the binary star population. Until then the absolute values should be treated with some caution.

It is emphasized that the data show a near perfect, though weak correlation with present-day density (fig.~4 in K12b). Thus, without more precise knowledge of the region's respective histories, there is no reason to argue that the dependence of MuF on density were too weak for consistency with the standard model.

\subsection{Representativity of star formation regions and their connection to the Galactic field}
\label{sec:field}
As clustered star formation is the dominant mode of star formation \citep{ll2003,porras2003}, it has been noted in the past that stars and binaries that make up the Galactic field population should originate from a clustered environment \citep{k95a,k95b,g2010,mk11}. Upon dissolution of such aggregates of stars, their members become part of the wider galaxy.

Based on this notion, K12b add up the binaries from star formation regions they investigated to compare this integrated population to solar-type field binaries \citep{r2010}. They thus implicitly assume that these seven regions, arbitrarily selected because we happen to have observational data for them, are representative of star formation environments that comprise nowadays the Galactic field.

K12b identify an overabundance of binaries between the integrated population and the field in group~(ii). They conclude that 'some regions \emph{must} underproduce close binaries, or the nearby young regions are not the source of most stars in the field'. As we do not know about the frequency a region of one kind contributed to the presently observed field population, we consider these conclusions to be too strong. If, e.g., regions of initial densities (not necessarily mass) closer to that of the ONC were the dominant donators, this would lead to lower binary proportions in these regions and would upon integrating also deplete the field binary population.

It is noteworthy, however, that Tau, Oph, Cha and USco individually exhibit an apparent excess of binaries near the peak of the solar-type field separation distribution (albeit with large statistical errors, Fig.~\ref{fig:obstheodata}) as also pointed out by K12b. For Tau and Cha, the theoretical distribution even lies well outside the observational error bars. If real, the origin remains elusive. Nevertheless, as pointed out before, the standard model is as of today in agreement with the data (including this excess), meaning that the hypothesis that the observed distribution has been selected from the theoretical one cannot be withdrawn with certainty. This finding does \emph{not} imply that the shown distribution is actually the true underlying parent distribution. A different one that better matches the binary excess and works as well might be invented. Future surveys covering smaller binary separations need to decide about the reality of this feature.

\section{Discussion}
\label{sec:universal}
In contrast to K12b who claim to rule out universal star formation, we here present further arguments in favour of the universality hypothesis of multiple star formation \citep{k2011,kp2011}. Whether or not star formation is a universal process is a key issue that affects a number of related fields.

Properties of binary populations do show correlations with (present-day) cluster parameters. \citet{s2007} show the MuF to anticorrelate with age in 13 low-density globular clusters, although this is not confirmed when adding clusters of higher density
\citep{m2012acs}. An even stronger anticorrelation with cluster luminosity is found for a larger sample by \citet{m2008}. A similar dependence on cluster mass is reported also for open clusters by \citet{s2010}. Recent results from the ACS survey confirm these trends \citep{m2012acs}. \citet{leigh2013} demonstrate one example of how this anticorrelation between cluster mass and (core) binary fraction can arise from a universal IBP.

For example, cataloging the absolute numbers and relative proportions of different types of stellar exotica (formed via binary internal evolution) is complementary to constraining the universality of star formation, both in clusters and the field.

The properties of some exotic populations thought to be descended from binary stars are also consistent with a universal IBP and MuF. For example, \citet{knigge2009} find the number of blue straggler (BS) stars in globular cluster cores to correlate sub-linearly with the core-mass ($N_{\rm BS}\propto$MuF$\times M_{\rm core}$). This is expected if all binaries have a binary (internal) evolution origin and cluster density increases with increasing cluster mass, which is the case for a weak initial mass-radius relation \citep[e.g.][]{mk12}. An analytical model for the formation of BS populations presented in \citet{leigh2011} requires an anticorrelation between MuF and cluster luminosity (a proxy for cluster mass) for better agreement with observed BS populations. This is naturally expected from the dynamical processing of binaries, which serves to destroy the binary progenitors of BSs within the context of the standard model. \citet{santana2013} recently considered BS population size as a function of different cluster properties, extending the pioneering work of \citet{piotto2004bs} to low-density environments. The authors demonstrated that BS population size is roughly constant below a critical cluster density, but decreases monotonically at higher densities. This can be interpreted as evidence that as you move to lower density environments, the hard--soft boundary is much wider than the separation corresponding to the BS progenitors. Thus, below some critical density, the BS populations should all be roughly the same fraction of the total cluster mass, which is consistent with what \citet{santana2013} see. In future studies, population sizes of these types of exotic populations can offer insight into constraining the initial cluster density in environments where it is difficult to measure orbital-parameter distributions.

If the initial distributions were \emph{wildly} different, why would one expect such behaviour for globular clusters and exotic populations?

\section{Summary}
\label{sec:summary}
Comparing multiplicity data from seven star-forming regions using a KS test, \citet*[K12b]{King2012b} found the respective separation distributions to be mostly indistinguishable. This has been confirmed here using a $\chi^2$-test although the detection rate of differences among the separation distributions in the investigated regions is a little larger.

K12b note the distribution's independence on the region's present-day number density which they assume not to be significantly different from the region's past. From this they draw the conclusion that stellar-density-dependent binary processing from the same universal IBP (the standard model) is incorrect in describing the observational data and that star formation cannot be universal.

In order to exclude density-dependent dynamical processing of binaries, it would need to be shown that dynamical processing is inconsistent with the data. This was not tested for in K12b. Already \citet{mk12} show that solutions to the observations exist within the standard model assuming initial densities were significantly larger. However, if initial densities were not much different from the presently observed densities, the standard model with the IBP used here would fail to reproduce some of the observations.

Here, we additionally demonstrated that neither KS nor $\chi^2$-test are suitable to detect differences among separation distributions which occur due to dynamical processing in the standard model. The same statistical results emerge when the standard model is assumed to be correct. These tests can thus not refute its validity. A larger number of binaries in the investigated separation ranges could enhance the detectability of differences among separation distributions when a $\chi^2$-test is applied (but not when a KS test is used since the separation shape alone is an insufficient tracer of dynamical processing).

We have further discussed K12b's assumption that the present-day density be a tracer of the initial density. While this may be put to discussion for individual regions, we make here the case that any region's stellar density should have been larger in the past. To be fair, the initial densities as constrained by \citet{mk12} and possibly required in the standard model are rather extreme and it is unclear whether or not these could have been this large. But these extreme constraints might currently as well be due to the statistical uncertainties affecting the observations. Unlike K12b, we further consider an age-dependent categorization of soft and hard binaries. In continuation, it is cautioned here \citep[as before in][]{mk12} that there is a degeneracy between age and initial density of a region in terms of its present multiplicity properties that may not be omitted. 

Finally, K12b's discussion about the Galactic field population makes the implicit assumption that the seven arbitrarily selected star formation regions for which observations happen to be available comprise a representative set from which the Galaxy field population originates. We consider this unlikely, but acknowledge an apparent excess of binaries near the peak of the field separation distribution in individual regions which are unexpected in, but do not rule out, the standard model. Instead when considering that now-dispersed clustered star formation regions followed a cluster mass spectrum, the \citet{mk11} binary population synthesis models have demonstrated that using the standard model the Galactic field population is reproduced upon adding up the contributions from clusters of different initial densities (both in terms of multiplicity fractions and distributions). These models, however, require star clusters to form compact initially.

In summary, after a re-analysis of the observed sample considered in K12b, we do not find convincing evidence against density-dependent binary processing, nor do we find evidence of non-universal star formation. Thus, the data are consistent with the universality hypothesis. Only if the densities had never been larger in the past would the standard model in its present formulation not account for all observations. Testing whether or not densities decreased to the present day might perhaps be possible from velocity dispersion measurements. A large value might indicate expansion which, given the youth of most regions, could result from binary processing of initially denser regions in the standard model. \textsc{Gaia} might help illuminating the issue. It is however mandatory to compare such measurements to dedicated computations of individual regions in order to exclude the possibility that observed velocity dispersions can occur in the standard model. In this respect, it might be worth investigating what parameters set the initial cluster density and/or size and mass \citep[see, e.g.,][]{mk10gc}. If we could understand what sets the initial density better, the results we present here show that this would contribute to answer questions regarding the universality of star formation.

\section*{Acknowledgements}
MM acknowledges support through DFG grant KR 1635/40-1. NL gratefully acknowledges the generous support of an NSERC Postdoctoral Fellowship. MG acknowledges a partial support by the National Science Centre through the grant
DEC-2012/07/B/ST9/04412. MM thanks Klaas S. de Boer and Ladislav~$\check{\rm S}$ubr for helpful discussions. MM also thanks Simon P. Goodwin for discussing their assumptions and providing the data used in K12b allowing a re-analysis. The authors thank an anonymous referee for a very helpful report which significantly contributed to improve the presentation of the manuscript.

\bibliographystyle{mn2e} \bibliography{biblio} \makeatletter  
\renewcommand{\@biblabel}[1]{[#1]}   \makeatother

\label{lastpage}

\end{document}